# How much is my car worth? A methodology for predicting used cars prices using  Random Forest

Nabarun Pal

Department of Metallurgical and Materials Engineering
Indian Institute of Technology Roorkee
Roorkee, India
pal@nabarun.in

Dhanasekar Sundararaman

Department of Information Technology
SSN College of Engineering
Chennai, India
dhanasekar312213@gmail.com

Priya Arora

Department of Computer Science
Texas A & M University
Texas, United States
arora.priya4172@gmail.com

Puneet Kohli

Department of Computer Science
Texas A & M University
Texas, United States
punkohl@gmail.com

Sai Sumanth Palakurthy

Department of Computer Science and Engineering
IIT (ISM) Dhanbad
Dhanbad, India
sumanth4591@gmail.com

*Abstract*—**Cars are being sold more than ever. Developing countries adopt the lease culture instead of buying a new car due to affordability. Therefore, the rise of used cars sales is exponentially increasing. Car sellers sometimes take advantage of this scenario by listing unrealistic prices owing to the demand. Therefore, arises a need for a model that can assign a price for a vehicle by evaluating its features taking the prices of other cars into consideration. In this paper, we use supervised learning method namely Random Forest to predict the prices of used cars. The model has been chosen after careful exploratory data analysis to determine the impact of each feature on price. A Random Forest with 500 Decision Trees were created to train the data. From experimental results, the training accuracy was found out to be 95.82%, and the testing accuracy was 83.63%. The model can predict the price of cars accurately by choosing the most correlated features.**

## I. INTRODUCTION

The prices of new cars in the industry is fixed by the manufacturer with some additional costs incurred by the Government in the form of taxes. So customers buying a new car can be assured of the money they invest to be worthy. But due to the increased price of new cars and the incapability of customers to buy new cars due to the lack of funds, used cars sales are on a global increase. Predicting the prices of used cars is an interesting and much-needed problem to be addressed. Customers can be widely exploited by fixing unrealistic prices for the used cars and many falls into this trap. Therefore, rises an absolute necessity of a used car price prediction system to effectively determine the worthiness of the car using a variety of features. Due to the adverse pricing of cars and the nomadic nature of people in developed countries, the cars are mostly bought on a lease basis, where there is an agreement between the buyer and seller. These cars upon completion of the agreement are resold. So reselling has become an essential part of today's world. Given the description of used cars, the prediction of used cars is not an easy task. There are a variety of features of a car like the age of the car, its make, the origin of the car (the original country of the manufacturer), its mileage (the number of kilometers it has run) and its horsepower. Due to rising fuel prices, fuel economy is also of prime importance. Other factors such as the type of fuel it uses, style, braking system, the volume of its cylinders (measured in cc), acceleration, the number of doors, safety index, size, weight, height, paint color,  consumer reviews, prestigious awards won by the car manufacturer. Other options such as sound system, air conditioner, power steering, cosmic wheels, GPS navigator all may influence the price as well. Some of the important features of this and their influence on price is detailed in Section III.

So, we propose a methodology using Machine Learning model namely random forest to predict the prices of used cars given the features. The price is estimated based on the number of features as mentioned above. The intricate details about this model on the used car's data set along with the accuracy are narrated in depth in Section V. We then deploy a website to display our results which are capable of predicting the price of a car given so many features of it. This deployed service is a result of our work, and it incorporates the data, ML model with the features.

To summarize,

- First, we collect the data about used cars, identify important features that reflect the price.
- Second, we preprocess and remove entries with NA values. Discard features that are not relevant for the prediction of the price.
- Third, we apply random forest model on the preprocessed dataset with features as inputs and the price as output.
- Finally, we deploy a web page as a service which incorporates all the features of the used cars and the random forest model to predict the price of a car.

The paper is organized as follows. Section II talks about the other works that predict the price of used cars. Section III details about the dataset and preprocessing. In Section IV we share the results of exploratory data analysis on our dataset. Section V talks about our proposed methodology of using random forests to predict car prices with details about the accuracy of training and test data. We conclude in Section VI with our future works in the last section.

## II. RELATED WORK

We use dataset from Kaggle for used car price prediction. The dataset contains various features as mentioned in section III of this paper that are required to predict and classify the range of prices of used cars. The literature survey provides few papers where researchers have used similar data set or related data-set for such price prediction.

[1] This patent describes a generic engine platform for assessing the price of an asset. This platform provides a price computation matrix for asset price prediction. To compute the price for vehicles, this platform may compute linear regression model that defines a set of input variables. However, it does not give details as what features can be used for specific type of vehicles for such prediction. We have taken important features for predicting the price of used cars using random forest models.

Zhang et al. [2] use Kaggle data-set to perform price prediction of a used car. The author evaluates the performance of several classification methods (logistic regression, SVM, decision tree, Extra Trees, AdaBoost, random forest) to assess the performance. Among all these models, random forest classifier proves to perform the best for their prediction task. This work uses five features (brand, powerPS, kilometer, sellingTime, VehicleAge) to perform the classification task after removal of irrelevant features and outliers from the dataset which gives an accuracy of 83.08% on the test data. We also use Kaggle data-set to perform prediction of used-car prices. However, the difference lies in the inclusion of few more relevant features in prediction model - the price of the car, and vehicleType. These two features play an important role in predicting the price of a used car which seems to be given less importance in the paper [2]. In addition to this, the range of features year of registration, PowerPS, the price seems to be narrowed down in work [2] due to which test data-set gives less accuracy w.r.t what we evaluate by broadening the range of the above-said features.

The report by Awad et al. [3] is more of an educational paper than a research paper. The author reviews six most popular classification methods (Bayesian classification, ANNs, SVMs, k-NN, Rough sets, and Artificial immune system) to perform a spam email classification task. The reason for choosing this paper is to understand these popular classification models in detail, and its applicability to the spam email classification problem since this paper gives much insight into each method. The main difference, however, between classifying price range and spam mail, is that spam email classification task is a binary one, whereas our motive is mainly one-vs-the-rest. The author uses Naive Bayes for classification which does not give accurate results due to its major concern of feature dependency as pointed out by the author. Due to this reason, we also did not try to evaluate the performance of our data-set using Naive Bayes model since our dataset has heavily feature dependency. To predict results with good accuracy, the author suggests a hybrid system which applies to our work by using Random Forest. A manipulation of various decorrelated decision trees, the Random Forest gives pretty good accuracy in comparison to prior work.

Work by Durgesh et al. [4] gives a good introductory paper on Support Vector Machine. The authors assess the performance of several classification techniques (K-NN, Rule-Based Classifiers, etc.) by performing the comparative assessment of SVM with others. This comparative study is done using several data-sets taken from the UCI Machine Learning Repository. This assessment yields that SVM gives much better classification accuracy in comparison to others. This gives us a baseline for prediction of tasks by using a simple linear model which gives good accuracy to let us use complex systems - random forest - which ultimately provides pretty good results for prediction of the used-cars price.

The Author of the paper [5] predicts the price of used cars in Mauritius by using four comparable machine learning algorithms - multiple linear regression, k-nearest neighbors, naive Bayes and decision trees algorithm. The author uses historical data collected from daily newspapers in Mauritius. The application of listed learning algorithms on this data provides comparable results with not-so-good prediction accuracy. The main difference, however, between classifying price range and spam mail, is that spam email classification task is a binary one, whereas our motive is mainly one-vs-the-rest. The difference between our analysis of this work is that we perform our assessment on data from Kaggle, whereas theirs is based on the data collected from the daily newspaper. In addition to this, the author uses simple and comparable classification algorithms that conform to our findings that using a sophisticated algorithm like random forest on our data-set can give pretty good results, and which has proved to be so.

Multiple regression models help in the classification of numerical values when there is more than one independent variable with one dependent variable. Noor et al. [6], hence, use this model to evaluate the prediction using https://www.pakwheels.com/ dataset. The authors use data-set of 2000 records collected within the duration of one or two months. The collected data includes features like color, advertisement date, etc. which seem to be not-so-relevant for such prediction, whereas, our model uses relevant features like the brand, kilometer, etc. which helps in predicting good accuracy using random forest on data-set obtained from Kaggle.

Researchers in the paper [7] use multivariate regression model in classification and prediction of used car prices. The authors use 2005 General Motor (GM) cars data-set for this classification task. They introduce variable selection techniques for determination of relevant features for such prediction tasks which provide insights of its applicability in several domains. The main emphasis of this paper is to discuss this model for learning and encouraging students to perform in this field. The model used in this work does not require any special knowledge of the dataset used. Hence the portal data (www.pakwheels.com) was sufficient to use. The difference, however, of their work with ours is that our work focusses upon preprocessing/filtering of data obtained from Kaggle with a selection of relevant features for prediction. This motivates us to generalize the model for a variety of brands with a range of years to predict prices using several relevant features.

## III. DATA SET AND PREPROCESSING

To accurately predict the prices of used cars, we used an open dataset to train our model. We used the 'Used Car Database' from Kaggle which is scraped from eBay-Kleinanzeigen, the German subsidiary of eBay, a publicly listed online classified portal. The dataset contains the prices and attributes of over 370,000 used cars sold on the website across 40 brands. Our dataset contains 20 unique attributes of a car being sold, out of which we removed a few irrelevant columns that have little to no impact on a car's price from our analysis, such as some pictures, postal code, and advertisement name. Additionally, we perform the following preprocessing steps on the data set helping us narrowing down the features –

1. Keep only listings for cars sold by private owners and filter out those sold by dealerships
2. Keep only listings for cars being sold, and filter out all request for purchase listings
3. Filter out cars manufactured before 1863 and after 2017, and derive the car's age
4. Filter out all cars with unrealistic Power values
5. Filter out listings which don't have an associated price
6. Filter out all cars listed as unavailable
7. Filter out invalid registration dates
8. Convert boolean (true/false) fields to numeric (0/1) based
9. Filter out all data with value as 'NA' (Not Available)

After pre-processing, the final dataset contains ten features for second hand cars – 'price', 'vehicleType', 'age', 'powerPS', 'model', 'kilometer', 'fuelType', 'brand', and 'damageRepaied', 'isAutomatic'. Out of these features, the most important for our prediction model is

1. price: The specified asking amount for the car
2. kilometer: A number of Kilometers the car has driven
3. brand: The car's manufacturing company
4. vehicleType: Whether a small car, limousine, bus, etc.

Just based on analyzing our input data set, we can see that on average, used cars were priced at approximately €11000 with an average kilometer reading of 125000 Km. 50% of cars from our dataset were being sold after 12 years of being used.

## IV. EXPLORATORY DATA ANALYSIS

After preprocessing the data, it is analyzed through visual exploration to gather insights about the model that can be applied to the data, understand the diversity in the data and the range of every field. We use a bar chart, box plot, distribution graph, etc. to explore each feature varies and its relation with other features including the target feature.

Figure 1 shows the distribution of age of the vehicle. The age is calculated using yearOfRegistration feature and current year (2017). Apparently, it is the number of years between the year of registration and 2017

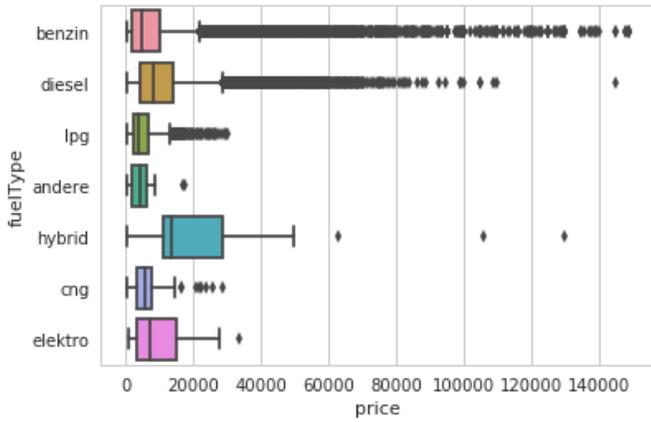

Fig. 1. Fuel Type vs. Price

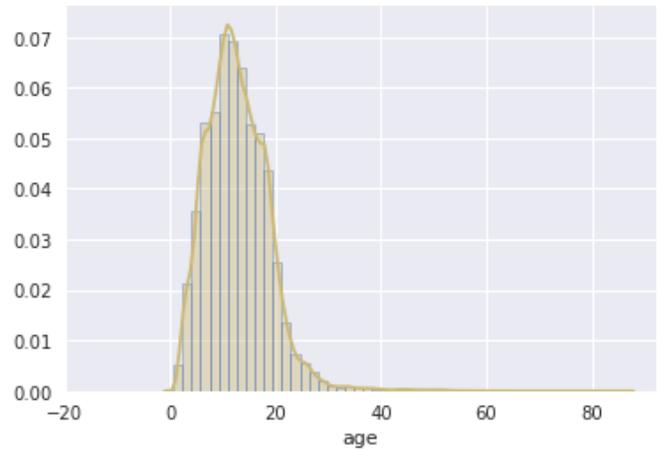

Fig. 3. Distribution of age of vehicle

Figure 2 shows the average prices of each vehicle type. It shows that vehicle types like suv, coupe, and cabrio have higher average prices. This information can be used to analyze the vehicle type people generally tend to buy.

Figure 4 shows the top ten average prices by brand. This bar graph shows that average price of Porsche is much higher, which is around 40000 while the next highest is around 20000

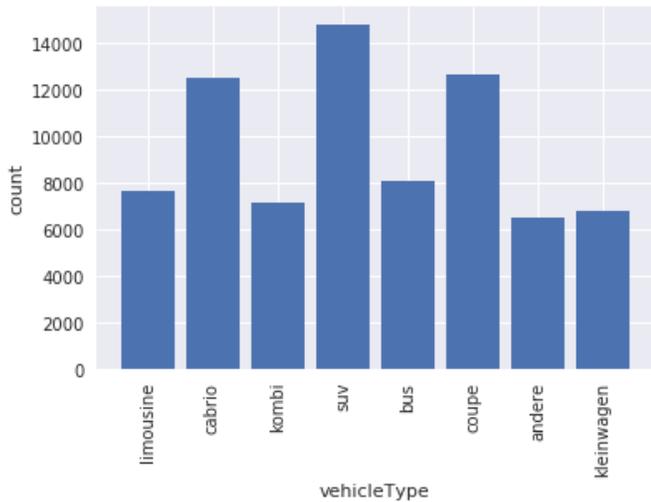

Fig. 2. Average Price for a specific vehicle type

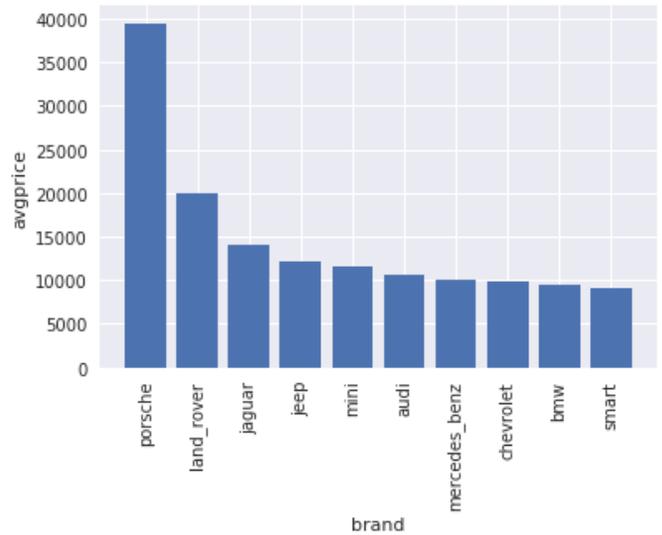

Fig. 4. Average Price for a specific brand of vehicle

Figure 3 shows the distribution of age of the vehicle. The age is calculated using yearOfRegistration feature and current year (2017). Apparently, it is the number of years between the year of registration and 2017. From the graph, it is evident that most of the vehicles are listed between 5 -15 years old. The graph provides us the insight that, people tend not to sell their vehicle neither too early, as it defeats the purpose of buying sometimes, nor too late as the worth of the car goes down dramatically.

Figure 5 shows that the average prices of these cars where the damage is repaired are higher than others. This may be obvious, but offers insights to the cost of a vehicle.

Figure 6 shows the number of entries having particular fuel type. From previous figures, it is evident that most of the cars have fuel type as benzin or diesel and they have a wide range of price distribution compared to the other fuel types.

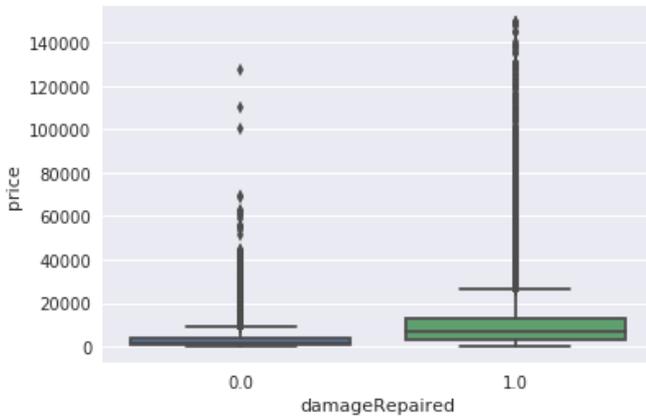

Fig. 5.   Box plot of damageRepaired vs. price

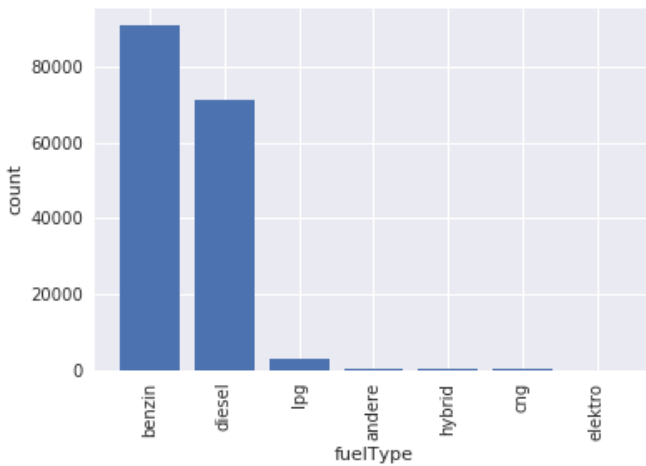

Fig. 6.   Counts per type of fuel type

## V.   MODEL DESCRIPTION

The problem at hand is a regression problem. We tried with linear regression and random forest regression. After much testing, it was found that random forest regression performed much better as it overcame the overfitting problem. The accuracy of the regression was less than 75% even in training data.

Random forest is primarily used for classification, but we used it as a regression model by converting the problem into an equivalent regression problem. Random forest comes under the category of ensemble learning methods, which contains a cluster of decision trees, usually hundreds or thousands in number. These trees are individually trained on parts of the dataset and help in learning highly unpredictable patterns by growing very deep. However, this may create an overfitting issue. This is overcome by averaging out the predictions of individual trees with a goal to reduce the variance and ensure consistency.

### A.   Model Parameters

Random Forest has several parameters to be tuned to which certain parameters have higher importance and are described below:

1.   Number of Estimators: This is the number of decision trees constituting the forest.

2.   A maximum number of features: It defines the maximum number of features a single decision tree should be trained.

A Grid Search Algorithm was employed to find the optimum number of trees, and best accuracy was found when 500 decision trees were used to build the forest. This was confirmed after iteratively increasing the number of decision trees in the multiples of 50.

Now, the maximum number of features is chosen to be equal to the number of features in the input data in case of regression problems and the square root of some features in case of classification. Since the problem at hand is a regression problem; we are going to the former.

### B.   Training and Testing

We split our input data into training, testing data and cross-validation with a 70:20:10 split ratio. The splitting was done by picking at random which results in a balance between the training data and testing data amongst the whole dataset. This is done to avoid overfitting and enhance generalization.

### C.   Accuracy

The model score is the coefficient of determination $R^2$ of the prediction. The training score was found out to be 95.82%, and the testing score was 83.63%.

The model was tuned in such a way that, only important features are taken and the rest are discarded. The important features are found using correlation, measuring their importance towards the estimation of the price of a vehicle. Overall, the random forest model effectively captured the nuances of the data and produced accurate predictions on the price of the vehicle.

## VI.   CONCLUSION

This paper evaluates used-car price prediction using Kaggle dataset which gives an accuracy of 83.62% for test data and 95% for train-data. The most relevant features used for this prediction are price, kilometer, brand, and vehicleType by filtering out outliers and irrelevant features of the dataset. Being a sophisticated model, Random Forest gives good accuracy in comparison to prior work using these datasets.

## VII.   FUTURE WORKS

Keeping the current model as a baseline, we intend to use some advanced techniques like fuzzy logic and genetic algorithms to predict car prices as our future work. We intend to develop a fully automatic, interactive system that contains a repository of used-cars with their prices. This enables a user to know the price of a similar car using a recommendation engine, which we would work in the future.